\documentclass[12pt,twoside]{article}
\usepackage{fleqn,espcrc1}

\input{psfig}
\usepackage{graphicx}
\usepackage[figuresright]{rotating}

\def\Journal#1#2#3#4{{#1} {\bf#2} (#4) #3}

\def\NPA{{\rm Nucl. Phys.} A}
\def\NPB{{\rm Nucl. Phys.} B}
\def\PLB{{\rm Phys. Lett.}  B}
\def\PRL{\rm Phys. Rev. Lett.}
\def\PRD{{\rm Phys. Rev.} D}
\def\PRC{{\rm Phys. Rev.} C}

\def\la{\langle}
\def\ra{\rangle}

\def\al{\alpha}

\def\be{\begin{equation}}
\def\ee{\end{equation}}
\def\bea{\begin{eqnarray}}
\def\eea{\end{eqnarray}}

\newcommand{\AmS}{{\protect\the\textfont2
  A\kern-.1667em\lower.5ex\hbox{M}\kern-.125emS}}

\begin{document}
\title{New effective treatment of the light-front nonvalence contribution
in timelike exclusive processes}

\author{Chueng-Ryong Ji\address[MCSD]{Department of Physics,
        North Carolina State University\\
        Raleigh, NC 27695-8202, USA}%
        \thanks{E-mail: crji@unity.ncsu.edu}
        and 
        Ho-Meoyng Choi\addressmark[MCSD]
         \address{Department of Physics, Carnegie Mellon University\\ 
           Pittsburgh, PA 15213, USA}
        \thanks{E-mail: homeoyng@andrew.cmu.edu}}


\maketitle

\begin{abstract}
We discuss a necessary nonvalence contribution in timelike exclusive 
processes. Following a Schwinger-Dyson type of approach, we relate the 
nonvalence contribution to an ordinary light-front wave function that has 
been extensively tested in the spacelike exclusive processes. 
A complicate four-body energy denominator is exactly cancelled in summing 
the light-front time-ordered amplitudes. Applying our method to 
$K_{\ell3}$ and $D^0\to K^- \ell^+ \nu_l$ where a rather substantial nonvalence 
contribution is expected, we find not only an improvement in comparing with
the experimental data but also a covariance(i.e. frame-independence) of 
existing light-front constituent quark model.
\end{abstract}

\baselineskip=20pt
With the wealth of new and upgraded B-meson factories, exclusive decay 
processes will be studied intensively. Unlike the leading twist 
structure functions measured in deep inelastic scattering, such exclusive
channels are sensitive to the structure of the hadrons at the 
amplitude level and to the coherence between the contributions of the 
various quark currents and multi-parton amplitudes. The central unknown 
required for reliable calculations of weak decay amplitudes are thus
the hadronic matrix elements. 

Perhaps, one of the most popular formulations for the analysis of
exclusive processes involving hadrons may be provided in the framework of 
light-front (LF) quantization~\cite{BPP}. 
In particular, the Drell-Yan-West ($q^+=q^0+q^3=0$) frame 
has been extensively used in the calculation
of various electroweak form factors and decay 
processes~\cite{Ja2,CJ1,Kaon}. As an example,
only the parton-number-conserving (valence) Fock state contribution 
is needed in $q^+=0$ frame when the ``good" component of the current,
$J^+$ or ${\bf J}_{\perp}=(J_x,J_y)$, is used for the spacelike 
electromagnetic form factor calculation of pseudoscalar mesons.
The LF approach may also provide a bridge between the two
fundamentally different pictures of hadronic matter, i.e. the 
constituent quark model (CQM) (or the quark parton model) closely
related to the experimental observations and the quantum chromodynamics 
(QCD) based on a covariant non-abelian quantum field theory.
The key of possible connection between the two pictures is the 
rational energy-momentum dispersion relation (i.e. the square root 
operator does not appear) that leads to a relatively simple
vacuum structure. There is no spontaneous creation of massive fermions
in the LF quantized vacuum. Thus, one can immediately obtain a
constituent-type picture, in which all partons in a hadronic state are
connected directly to the hadron instead of being simply disconnected 
excitations (or vacuum fluctuations) in a complicated medium.
A possible realization of chiral symmetry breaking in the LF vacuum
has also been discussed in the literature~\cite{Wilson}.

On the other hand, the analysis of timelike exclusive processes 
has remained
as a rather significant challenge in the LF approach. In principle, the 
$q^+\neq0$ frame can be used to compute the timelike processes but 
then it is inevitable to encounter the particle-number-nonconserving 
Fock state (or nonvalence) contribution.
The main source of difficulty in CQM phenomenology 
is the lack of information on the non-wave-function vertex(black blob 
in Fig.~1(a)) in the nonvalence diagram arising from the quark-antiquark 
pair creation/annihilation. The non-wave-function vertex(black blob) was 
recently also called the embedded state~\cite{BCJ}. 
This should contrast with the white blob representing the usual LF 
valence wave function. 
\begin{figure}[t]
\centerline{\psfig{figure=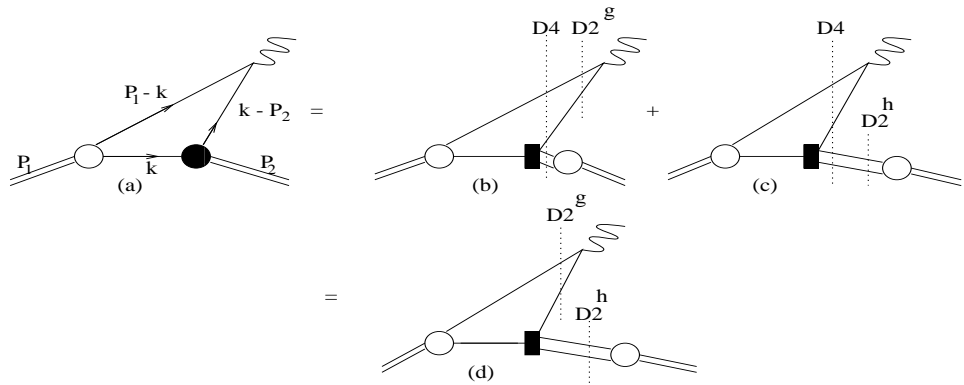,height=2in,width=5in}}
\caption{Effective treatment of the LF nonvalence amplitude.}
\end{figure}
In principle, there is a systematic program
laid out by Brodsky and Hwang~\cite{BH} to include the 
particle-number-nonconserving
amplitude to take into account the nonvalence contributions.
However, the program requires to find all the higher Fock-state wave 
functions while there has been relatively little progress in computing 
the basic wave functions of hadrons from first principles. 
Recently, a method of analytic continuation from the spacelike region has
also been suggested to generate necessary informations in the timelike
region without encountering a direct calculation of the nonvalence 
contribution~\cite{Anal}. 
Even though some explicit example has been presented
for manifestly covariant theoretical models, 
this method has not yet been implemented to more realistic 
phenomenological models. 

In this letter, we thus present an alternative way of handling the nonvalence
contribution. Our aim of new treatment is to make the program more
suitable for the CQM phenomenology specific to the low momentum transfer 
processes. Incidentally, the light-to-light ($K_{\ell3}$) and heavy-to-light 
($D^0\to K^-\ell^+\nu_\ell$) decays involving rather low momentum transfers 
bear a substantial contribution from the nonvalence part and their 
experimental data are better known than other semileptonic processes with 
large momentum transfers.
Including the nonvalence contribution, our results on 
$K_{\ell3}$ and $D^0\to K^-\ell^+\nu_\ell$ not only show a definite 
improvement in comparison with experimental data but also exhibit a 
covariance (i.e frame-independence) of our approach.

The semileptonic decay of $Q_{1}\bar{q}$ bound state with four-momentum
$P^\mu_1$ and mass $M_1$ into another $Q_{2}\bar{q}$ bound state with 
$P^\mu_2$ and $M_2$ is governed by the weak current, viz.,
\be{\label{eq:K1}}
J^{\mu}(0)=\la P_{2}|\bar{Q_{2}}\gamma^{\mu}Q_{1}|P_{1}\ra
= f_{+}(q^{2})(P_{1}+P_{2})^{\mu} + f_{-}(q^{2})q^{\mu},
\ee
where $q^\mu=(P_{1}-P_{2})^\mu$ is the four-momentum transfer to
the lepton pair ($\ell\nu$) and $m^{2}_\ell\leq q^2\leq (M_{1}-M_{2})^{2}$.
The covariant three-point Bethe-Salpeter (BS) amplitude of the total current 
$J^\mu(0)$ in Eq.~(\ref{eq:K1}) may be given by \bea{\label{eq:Jmu}}
J^{\mu}(0)&=&iN_c\int \frac{d^{4}k}{(2\pi)^4}
\frac{H^{\rm cov}_1H^{\rm cov}_2S^\mu}{(p^{2}_{1}-m^{2}_{1}+i\epsilon)
(p^{2}_{2}-m^{2}_{2}+i\epsilon)}
\frac{1}{(p^{2}_{\bar{q}}-m^{2}_{\bar{q}}+i\epsilon)},
\eea
where $N_c$ is the color factor, $H^{\rm cov}_{1[2]}$ is
the covariant intitial[final] state meson-quark vertex function
that satisfies the BS equation, and 
$S^\mu = {\rm Tr}[\gamma_{5}({\not\! p}_{1}+m_{1})\gamma^\mu
({\not\! p}_{2}+m_{2})\gamma_{5} (-{\not\!p}_{\bar{q}}+m_{\bar{q}})]$.
The quark momentum variables are given by $p_1=P_1 - k$, $p_2=P_2-k$, and 
$p_{\bar{q}}=k$. 
As shown in the literature~\cite{BCJ}, the LF energy integration reveals 
an explicit correspondence between the sum of LF time-ordered 
amplitudes and the original covariant amplitude.
For instance, performing the $k^-$ pole integration, 
we obtain the LF currents, $J^\mu_V$ and $J^\mu_{NV}$ corresponding to 
the usual LF valence diagram and the nonvalence diagram shown in 
Fig.~1(a), respectively. 
Since $H^{\rm cov}_{2}$ satisfies the BS equation, we iterate 
$H^{\rm cov}_{2}$ once and perform its LF energy integration to find the 
corresponding LF time-ordered diagrams Figs.~1(b) and~1(c) after the iteration.
The similar idea of iteration in a Schwinger-Dyson (SD) type of approach 
was presented in Ref.~\cite{BJS} to pin down the LF bound-state equation
starting from the covariant BS equation. 
Comparing the LF time-ordered expansions before and after the
iteration, we realize that the following link between the 
non-wave-function vertex (black blob) and the 
ordinary LF wave function (white blob) as shown in Fig.~2
naturally arises, i.e., 
\bea\label{eq:SD}
(M^2-M'^{2}_0)\Psi'(x_i,{\bf k}_{\perp i})
=\int[dy][d^2{\bf l}_{\perp}]
{\cal K}(x_i,{\bf k}_{\perp i};y_j,{\bf l}_{\perp j})
\Psi(y_j,{\bf l}_{\perp j}),
\eea 
where $M$ is the mass of outgoing meson and $M'^{2}_0=(m^2_1+{\bf 
k}^2_{\perp 1})/x_1 - (m^2_2+{\bf k}^2_{\perp 2})/(-x_2)$ with
$x_1 = 1-x_2 > 1$ due to the kinematics of the non-wave-function vertex.
\begin{figure}[t]
\centerline{\psfig{figure=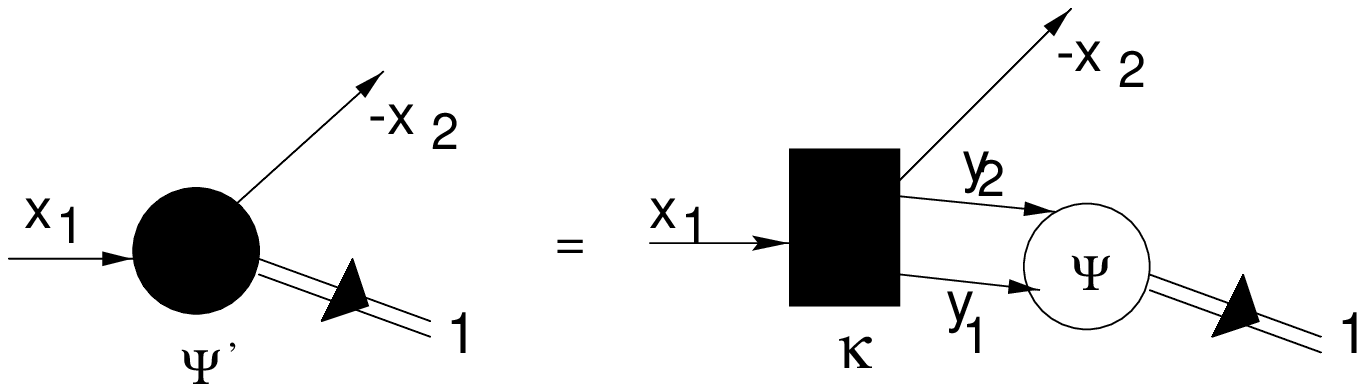,height=1.0in,width=3.2in}}
\caption{Non-wave-function vertex(black blob) linked to an ordinary
LF wave function(white blob).}
\end{figure}
We note that Eq.~(\ref{eq:SD}) essentially takes the same form as 
the LF bound-state equation (similar to the LF projection of BS 
equation) except the difference in kinematics(e.g. $-x_2 >0$ for the 
non-wave-function vertex).
Incidentally, Einhorn~\cite{Einhorn} also discussed the extension of the 
LF BS amplitude in 1+1 QCD to a non-wave-function vertex similar to 
what we obtained in this work. 

In the above procedure, we also find that the 
four-body energy denominator $D_4$ is exactly cancelled in the sum of LF 
time-ordered amplitudes as shown in Figs.~1(b) and~1(c), i.e., 
$1/D_4D^g_2 + 1/D_4D^h_2 =1/D^g_2D^h_2$.
We thus obtain the amplitude identical to the nonvalence
contribution in terms of ordinary LF wave functions of gauge boson($W$) 
and hadron (white blob) as drawn in Fig.1(d). This method, however,
requires to have some relevant operator depicted as the black 
square(${\cal K}$) in Fig.~2(See also Fig.1(d)), that is in general 
dependent on the involved momenta connecting one-body to three-body sector.
We now present some details of kinematics in the semileptonic decay 
processes to discuss a reasoning of how we handle the nonvalence 
contribution involving the momentum-dependent ${\cal K}$ 
for relatively small momentum transfer processes such 
as $\pi_{e3}$, $K_{\ell3}$ and $D\to K\ell\nu$. 

In purely longitudinal momentum 
frame where $q^+>0$ and ${\bf P}_{1\perp}={\bf P}_{2\perp}=0$,  
the momentum transfer $q^2=q^+q^-$ 
can be written in terms of the momentum fraction 
$\alpha=P^{+}_{2}/P^{+}_{1}=1-q^{+}/P^{+}_{1}$
as $q^{2}=(1-\alpha)(M^{2}_{1}-M^{2}_{2}/\al)$.
With the iteration procedure in this frame,
the results for the ``$+$"-component of the current 
$J^\mu$ are given by
\be{\label{eq:JV}}
J^{+}_{\rm V}=\frac{N_c}{16\pi^3}
\int^{\alpha}_{0}dx\int d^{2}{\bf k}_{\perp}
\frac{\Psi_{i}(x,{\bf k}_{\perp}) S^{+}_{V}
\Psi_{f}(x',{\bf k}_{\perp})}{x(1-x)(1-x')},
\ee
and
\bea{\label{eq:JNV}}
J^{+}_{NV}&=&\frac{N_c}{16\pi^3}
\int^{1}_{\alpha}dx
\int d^{2}{\bf k}_{\perp}\frac{\Psi_i(x,{\bf k}_{\perp})S^{+}_{NV}}
{x(1-x)(x'-1)}\biggl(\frac{1}{\alpha D^g_2}\biggr)\nonumber\\
&&\times\int\frac{dy}{y(1-y)}\int d^2{\bf l}_{\perp}
{\cal K}(x,{\bf k}_{\perp};y,{\bf l}_{\perp})\Psi_{f}(y,{\bf l}_{\perp}),
\eea
where $x=k^+/P^{+}_{1}$, $x'=x/\alpha$, $D^g_{2}=q^- - p^{-}_{1}-(-p^{-}_{2})$
with $-p^{-}_{2}>0$ and $\Psi_{i[f]}(x,{\bf k}_{\perp})
=h^{\rm LF}_{1[2]}/(M^{2}_{1[2]}-M^{2}_{01[02]})$ with
$M^{2}_{01}[M^2_{02}]=m^{2}_{1\perp}/(1-x)
+ m^{2}_{\bar{q}\perp}/x [m^{2}_{2\perp}/(1-x')+m^{2}_{\bar{q}\perp}/x'$]
and $m^2_{i\perp}=m^2_i+{\bf k}^2_{\perp}$.
Here, Eq.~(\ref{eq:SD})(SD type equation) was
folded in the derivation of Eq.~(\ref{eq:JNV}) by the iteration procedure.
While the relevant operator ${\cal K}$ is in general 
dependent on all internal momenta ($x,{\bf k}_{\perp},y,{\bf l}_{\perp}$), 
a sort of average on ${\cal K}$ over $y$ and ${\bf l}_{\perp}$
in Eq.(5) which we define as 
$G_{P_1P_2}\equiv\int[dy][d^2{\bf l}_{\perp}]
{\cal K}(x,{\bf k}_\perp;y,{\bf l}_\perp)\Psi_f(y,{\bf l}_{\perp})$
is dependent only on $x$ and ${\bf k}_{\perp}$. 
Now, the range of the momentum fraction $x$ depends on the 
external momenta for the embedded states. 
As shown in Eq.(5), the lower bound of $x$ 
for the kernel in the nonvalence contribution is given by $\alpha$ which 
has the value $\alpha=M_2/M_1$ at the maximum $q^2$.
As the mass difference between the primary and secondary mesons gets 
smaller, not only the range of $q^2$ is reduced but also $\alpha$ gets 
closer to 1. Perhaps, the best experimental process for such limit
may be the pion beta decay $\pi^\pm \to \pi^0 e^\pm {\bar \nu_e}$, where
our numerical prediction $f_-(0)/f_+(0)= -3.2\times 10^{-3}$
following the treatment presented in this work is in an excellent
agreement with $-3.5\times 10^{-3}$ obtained by the 
method proposed by Jaus including the zero-modes~\cite{Ja1}.
In Ademollo-Gatto's SU(3) limit~\cite{AG}, the $q^2$ range of 
the nonvalence contribution shrinks to zero and $\alpha$ becomes 
precisely 1. 
However, even if $\alpha$ is not so close to 1, the initial wavefunction 
$\Psi_i(x,{\bf k}_{\perp})$ plays the 
role of a weighting factor in the nonvalence contribution and enfeeble 
the contribution from the region of $x$ near 1. Thus, 
for the processes that we discussed in this letter,
the effective $x$ region for the nonvalence contribution
is quite narrow. Similarly, the region of the 
transverse momentum ${\bf k}_{\perp}$ is also limited only up to the 
scale of hadron size due 
to the same weighting factor $\Psi_i(x,{\bf k}_{\perp})$. In this work, 
we thus approximate $G_{P_1P_2}$ as a constant and examine the validity 
of this approximation by checking the frame independence of our 
numerical results. 

In Eqs.(4) and (5), the trace terms $S^+_{V}(p^{-}_{\bar{q}}=k^{-}_{\rm on})
=(4P^{+}_{1}/x')\{ {\bf k}^{2}_{\perp}
+ [x m_1  + (1-x)m_{\bar{q}}][x'm_2 + (1-x')m_{\bar{q}}]\}$ 
and $S^{+}_{NV}= S^{+}_{V}(p^-_i=p^-_{i{\rm on}})
+ 4p^{+}_{\rm 1on}p^{+}_{\rm 2on}(p^{-}_{\bar{q}} 
- p^{-}_{\bar{q}{\rm on}})$ correspond to the product of initial and
final LF spin-orbit wave functions that are uniquely determined by a
generalized off-energy-shell Melosh transformation. Here,  
the subscript (on) means on-mass-shell and the instantaneous part of 
nonvalence diagram corresponds to 
$S^+_{NV}-S^+_{V}(p^-_{i}=p^-_{i{\rm on}})$.
While the LF vertex function $h^{LF}_{1[2]}$ formally stems from
$H^{\rm cov}_{1[2]}$, practical informations on the radial wave function
$\Psi_{i[f]}(x,{\bf k}_{\perp})$ (consequently $h^{LF}_{1[2]}$) can be
obtained by LF CQM. The details of our variational procedure to determine
both mass spectra and wave functions of pseudoscalar mesons were recently
documented in Refs.~\cite{CJ1,Kaon} along with an extensive test of the model
in the spacelike exclusive processes. The same model is used in this work.

For the check of frame-independence, 
we also compute the ``$+$" component of the current $J^\mu_{D}$ 
in the Drell-Yan-West ($q^+=0$) frame where only valence contribution 
exists. 
Since the form factor $f_+(q^2)$ obtained
from $J^+_D$ in $q^+=0$ frame is immune to the zero-mode 
contribution~\cite{Kaon,BH,Chang,Ja1}, 
the comparison of $f_+(q^2)$ in the two completely different frames
(i.e. $q^+=0$ and $q^+\neq0$) would reveal the validity of existing model
with respect to a covariance. 
The comparison of $f_-(q^2)$, however, cannot give a meaningful test
of covariance because of the zero-mode complication as noted in
Ref.~\cite{Ja1}. Indeed, the difference between the two ($q^+=0$ and
$q^+\neq0$) results of $f_-(q^2)$ amounts to the zero-mode 
contribution~\cite{Chang}.

\begin{figure}[t]
\centerline{\psfig{figure=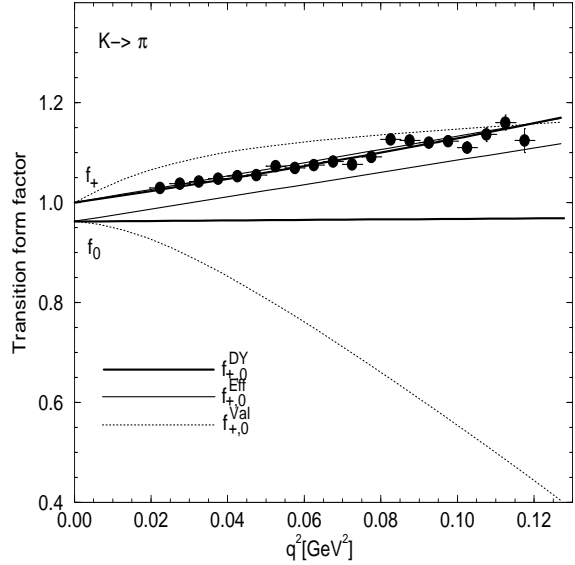,height=3.5in,width=3.5in}}
\caption{The weak form factors for
$K^{0}_{\ell3}$ compared with the experimental data~\protect\cite{Apo}.}
\end{figure}
In our numerical calculation for the processes of $K_{\ell3}$ and 
$D^0\to K^-\ell^+\nu$ decays, we use the linear potential parameters presented
in Ref.~\cite{Kaon}. In Fig.~3, we show the weak form factors $f_+(q^2)$ and 
$f_0(q^2)=f_+(q^2)+q^2 f_-(q^2)/(M^2_1-M^2_2)$ for 
$K^{0}_{\ell3}$ decays. 
The thick solid lines are our analytic solutions obtained from 
the $q^+=0$ frame; note here again that the lower thick solid line
($f_0$) in Fig.~3 is only the partial result without including the
zero-mode contribution while the upper thick solid line 
($f_+$ immune to the zero-mode) is the full result. 
The thin solid lines are the full results of our effective calculations 
with a constant ($G_{K\pi}$=3.95) fixed by the normalization of $f_+$ at 
$q^2=0$ limit.
For comparison, we also show only the valence 
contributions(dotted lines) in $q^+\neq0$ frame.
As expected, a clearly distinguishable nonvalence contribution is found.
Following the popular linear parametrization~\cite{Data}, we 
plot the results of our effective solutions(thin solid lines) using
$f_{i}(q^2)=f_{i}(q^2=m^{2}_{\ell})(1+\lambda_{i}q^2/M^{2}_{\pi^+})(i=+,0)$.
In comparison with the data, the same normalization as the data
$f_{+}(0)=1$~\cite{Apo} was used in Fig.~3.
Our effective solution(upper thin solid line) is not only 
in a good agreement with the data~\cite{Apo}
but also almost identical to that in $q^+=0$ frame(upper thick 
solid line) indicating the frame-independence of our model. 
Note also that the difference in $f_0(q^2)$ between $q^+\neq0$(lower 
thin solid line) and $q^+=0$(lower thick solid line) frames amounts 
to the zero-mode contribution~\cite{Chang}. 

\begin{table}[t]
\caption{ Model predictions for the parameters of $K^{0}_{\ell3}$ decays.
The decay width is in units of $10^{6}$ s$^{-1}$. The
used CKM matrix is $|V_{us}|=0.2196\pm0.0023$~\protect\cite{Data}.}
\begin{center}
\begin{tabular}{|c|c|c|c|}
\hline
&Effective & $q^+=0$ & Experiment\\
\hline
$f_{+}(0)$ & 0.962 [0.962] & 0.962 [0.962] & \\
\hline
$\lambda_{+}$& 0.026 [0.083] & 0.026 [0.026] & $0.0288\pm0.0015
[K^{0}_{e3}]$\\
\hline
$\lambda_{0}$& 0.025 [$-0.017$]& 0.001 [$-0.009$]
& $0.025\pm0.006[K^{0}_{\mu3}]$\\
\hline
$\xi_{A}$& $-0.013$ [$-1.10$]& $-0.29 [-0.41]$
& $-0.11\pm0.09[K^{0}_{\mu3}]$\\
\hline
$\Gamma(K^{0}_{e3})$ & $7.3\pm0.15$
&  $7.3\pm0.15$ & 7.5$\pm$0.08\\
\hline
$\Gamma(K^{0}_{\mu3})$ & $4.92\pm0.10$
&  $4.66\pm0.10$ & 5.25$\pm$0.07\\
\hline
\end{tabular}
\end{center}
\end{table}
In comparison with experimental data, we summarized our results of 
several experimental observables in Table~1;
i.e. the actual value of $f_+(0)$, the slopes $\lambda_+$  
[$\lambda_0$] of $f_+(q^2)$ [$f_0(q^2)$] at $q^2=0$, 
$\xi_{A}$=$f_-(0)/f_+(0)$, and the decay rates $\Gamma(K^{0}_{e3})$
and $\Gamma(K^{0}_{\mu3})$. 
In the second column of Table~1,
our full results including nonvalence contributions are presented along 
with the valence contributions in the square brackets. In the third 
column of Table~1, the results in $q^+=0$ frame are presented with[without] 
the instantaneous part. As one can see in Table 1, adding the nonvalence
contributions clearly improves the results of $\lambda_0$, i.e.
our full result of $\lambda_0$= 0.025 is in an excellent agreement with 
the data, $\lambda^{\rm Exp.}_0$= 0.025$\pm$0.006.
The results of $\xi_{A}$ and $\Gamma(K^{0}_{\mu3})$ seem to be also 
improved by our full effective calculations including the nonvalence
contributions.

In Figs.~4(a,b), we show the weak form factors for $D^0\to K^-\ell^+\nu$ 
decays and compare with the experimental data~\cite{Data}(full dot) with an 
error bar at $Q^2=0$ as well as the lattice QCD results~\cite{Ber}(circle 
and square) 
and~\cite{Bow}(cross). All the line assignments are same as in Fig.~3.
In Fig.~4(a), the thin solid line of our full result in $q^+\neq0$ is not 
visible because it exactly coincides with the thick solid line of
the result in $q^+ = 0$ confirming the frame-independence of our 
calculations. Our value of $f_+(0)$=0.736 is also within the error bar of 
the data~\cite{Data}, $f^{\rm Exp.}_+(0)$=0.7$\pm$0.1. In Fig.~4(b),
the difference between the thin and thick solid lines is the measure
of the zero-mode contribution to $f_0(q^2)$ in $q^+=0$ frame~\cite{Chang}.
The form factors obtained from our effective 
calculations($G_{DK}$=3.5) are also plotted with the usual 
parametrization of pole dominance model, i.e. 
$f_{+(0)}(q^2)=f_{+(0)}(0)/(1-q^2/M^2_{1^{-}(0^+)})$.
Our pole masses turn out to be $M_{1^-}$=2.16 GeV and $M_{0^+}$=2.79 GeV, 
respectively, and we note that $M_{1^-}$=2.16 GeV is in a good 
agreement with the mass of $D^*_{s}$,i.e. 2.1 GeV. 
Using CKM matrix element $|V_{cs}|=1.04\pm 0.16$~\cite{Data},
our branching ratios ${\rm Br}(D^{0}_{e3}) = 3.73\pm 1.24$ and
${\rm Br}(D^{0}_{\mu3}) = 3.60\pm 1.19$ are also comparable with the
experimental data $3.64\pm 0.18$ and $3.22\pm 0.17$~\cite{Data}, respectively.
\begin{figure}
\centerline{\psfig{file=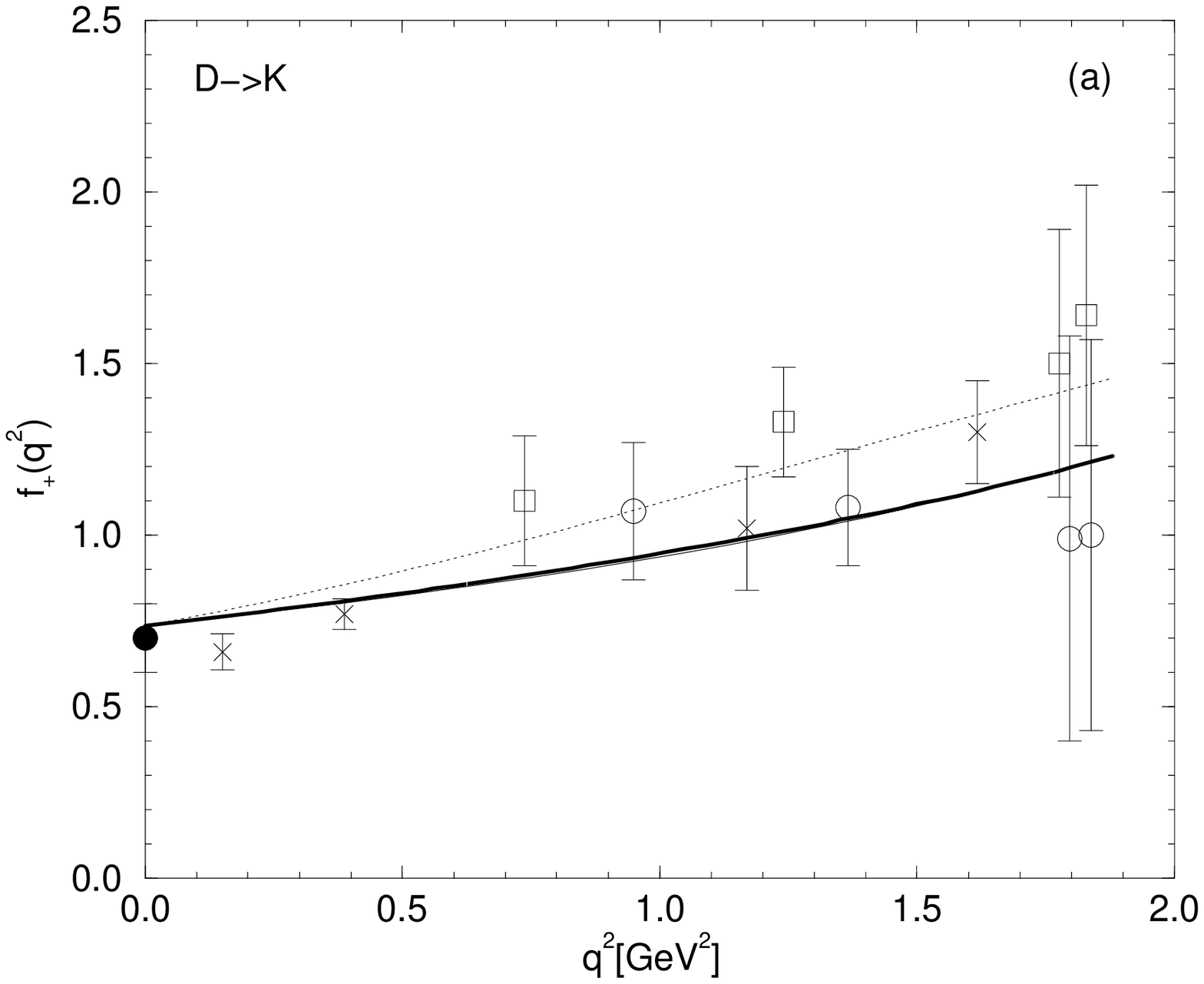,height=3.5in,width=3.5in}}
\centerline{\psfig{file=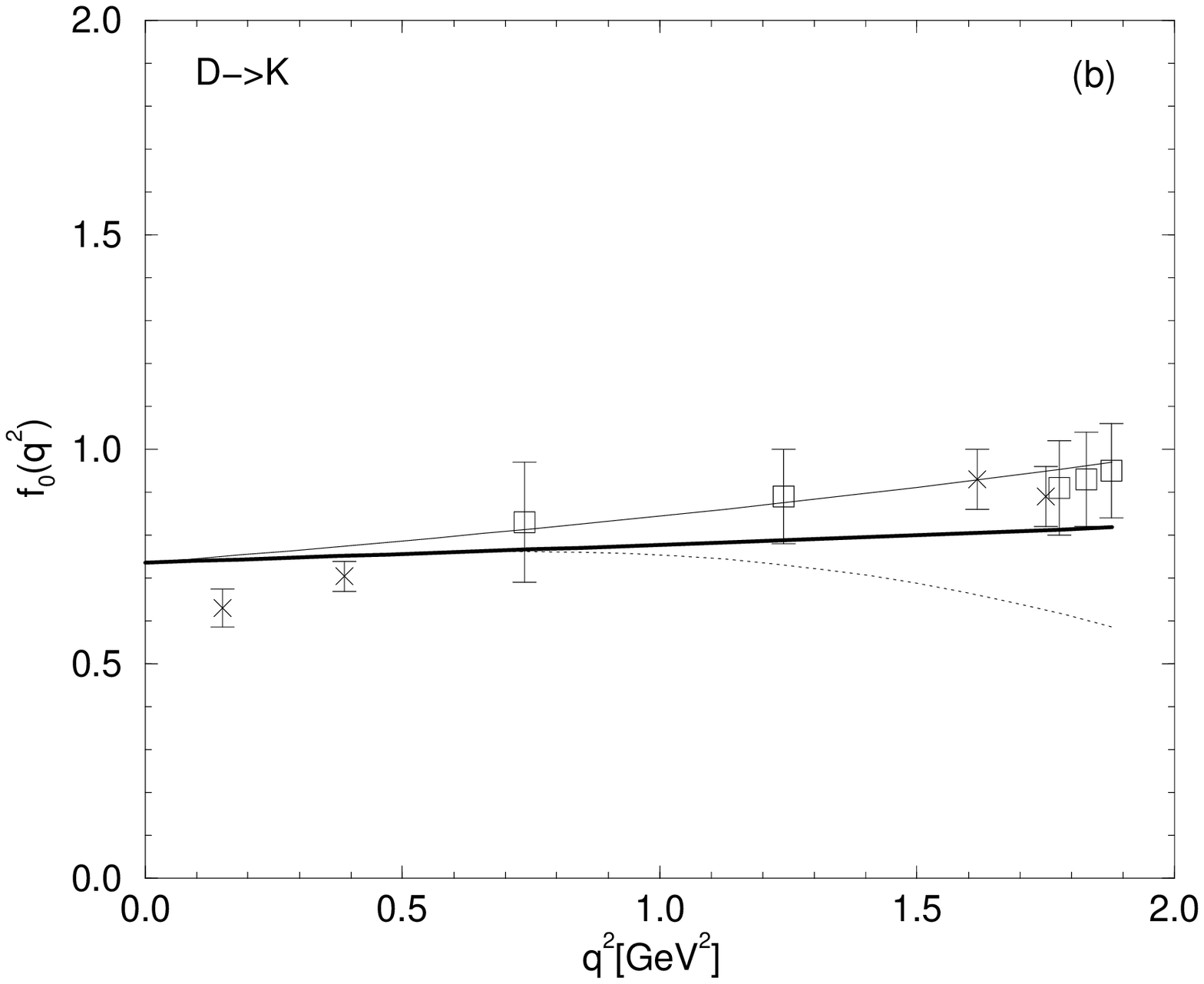,height=3.5in,width=3.5in}}
\caption{The weak form factors for $D\to K$
transition.}
\end{figure}

In summary, we presented an effective treatment of the LF
nonvalence contributions crucial in the timelike exclusive processes.
Using a SD-type approach and summing the LF time-ordered amplitudes,  
we obtained the nonvalence contributions in terms of ordinary LF 
wavefunctions of gauge boson and hadron
that have been extensively tested in the spacelike exclusive processes.  
Including the nonvalence contribution, our results show a definite 
improvement in comparison with experimental data on $K_{\ell3}$
and $D^0\to K^-\ell^+\nu_\ell$ decays. Our result on $\pi_{e3}$
is also consistent with the result obtained by other method.
Furthermore, the frame-independence of our 
results indicate that a constant $G_{P_1 P_2}$ is an approximation 
appropriate to the small momentum transfer processes. Applications
to the heavy-to-light decay processes involving large 
momentum transfers would require an improvement on 
this approximation perhaps guided by the perturbative QCD approach.
Consideration along this line is underway.

This work was supported by the US DOE under contracts DE-FG02-96ER40947.
The North Carolina Supercomputing Center and the National Energy Research
Scientific Computer Center are also acknowledged for the grant of 
supercomputer time.

\end{document}